\documentclass{jfm}

\usepackage{graphicx}
\usepackage{natbib}
\usepackage{tikz,amsmath,amssymb,esint,dcolumn}

\ifCUPmtlplainloaded \else
  \checkfont{eurm10}
  \iffontfound
    \IfFileExists{upmath.sty}
      {\typeout{^^JFound AMS Euler Roman fonts on the system,
                   using the 'upmath' package.^^J}%
       \usepackage{upmath}}
      {\typeout{^^JFound AMS Euler Roman fonts on the system, but you
                   dont seem to have the}%
       \typeout{'upmath' package installed. JFM.cls can take advantage
                 of these fonts,^^Jif you use 'upmath' package.^^J}%
      }
  \else
  \fi
\fi

\ifCUPmtlplainloaded \else
  \checkfont{msam10}
  \iffontfound
    \IfFileExists{amssymb.sty}
      {\typeout{^^JFound AMS Symbol fonts on the system, using the
                'amssymb' package.^^J}%
       \usepackage{amssymb}%
       \let\le=\leqslant  
       \let\ge=\geqslant  
      }{}
  \fi
\fi

\ifCUPmtlplainloaded \else
  \IfFileExists{amsbsy.sty}
    {\typeout{^^JFound the 'amsbsy' package on the system, using it.^^J}%
     \usepackage{amsbsy}}
    {\providecommand\boldsymbol[1]{\mbox{\boldmath $##1$}}}
\fi

\newcommand\solidrule[1][21pt]{\rule[0.5ex]{#1}{.4pt}}

\newcommand\dottedrule{\mbox{%
	\solidrule[2pt]\hspace{3pt}\solidrule[2pt]\hspace{3pt}\solidrule[2pt]\hspace{3pt}\solidrule[2pt]\hspace{3pt}\solidrule[2pt]}}

\newcommand{\bu}{\mathbf{u}}
\newcommand{\bx}{\mathbf{x}}
\newcommand{\bk}{\mathbf{k}}
\providecommand\bnabla{\boldsymbol{\nabla}}
\providecommand\bcdot{\boldsymbol{\cdot}}
\newcommand{\wT}{\lan wT \ran}
\newcommand{\T}{\lan T-\oT_T \ran}
\newcommand{\oT}{\overline{T}}
\newcommand{\owT}{\overline{wT}}
\newcommand{\oJ}{\overline{J}}
\newcommand{\DT}{\delta \oT}

\newcommand{\cQ}{\mathcal Q}
\newcommand{\cQk}{\mathcal Q_\bk}
\newcommand{\lan}{\left\langle}
\newcommand{\ran}{\right\rangle}

\title[Internally heated convection beneath a poor conductor]{Internally heated convection\\beneath a poor conductor}

\author[D.\ Goluskin]{David Goluskin$^1$%
  \thanks{Email address for correspondence: goluskin@umich.edu}}

\affiliation{$^1$Mathematics Department, University of Michigan, Ann Arbor, MI 48109, USA}

\pubyear{?}
\volume{?}
\pagerange{?}
\date{?; revised ?; accepted ?. - To be entered by editorial office}

\begin{document}

\maketitle

\begin{abstract}
We consider convection in an internally heated layer of fluid that is bounded below by a perfect insulator and above by a poor conductor. The poorly conducting boundary is modelled by a fixed heat flux. Using solely analytical methods, we find linear and energy stability thresholds for the static state, and we construct a lower bound on the mean temperature that applies to all flows. The linear stability analysis yields a Rayleigh number above which the static state is linearly unstable ($R_L$), and the energy analysis yields a Rayleigh number below which it is globally stable ($R_E$). For various boundary conditions on the velocity, exact expressions for $R_L$ and $R_E$ are found using long-wavelength asymptotics. Each $R_E$ is strictly smaller than the corresponding $R_L$ but is within 1\%. The lower bound on the mean temperature is proven for no-slip velocity boundary conditions using the background method. The bound guarantees that the mean temperature of the fluid, relative to that of the top boundary, grows with the heating rate ($H$) no slower than $H^{2/3}$.
\end{abstract}


\section{Introduction}
\label{sec: intro}

Mathematical models of thermal convection in horizontal fluid layers are studied both as examples of complexity in nonlinear systems and as idealizations of convention in astrophysical, geophysical, and engineering applications. Convection in a layer can be driven by internal heating or cooling, by the boundary conditions, or both. Rayleigh--B\'enard (RB) convection \citep{Siggia1994, Getling1998, Ahlers2009}, which has enjoyed the most attention, is driven solely by the boundary conditions. Internally heated (IH) convection, which is no less fundamental, is driven in its simplest models by constant and uniform volumetric heating. The IH configuration most commonly studied is a fluid layer bounded below by a perfect insulator and above by a perfect conductor. Here, we study a model of IH convection that also is bounded below by a perfect insulator but is bounded above by a \emph{poor} conductor -- a configuration considered in very few previous works \citep{Hewitt1980, Ishiwatari1994}. This model also describes the dynamics of internally \emph{cooled} convection with the top and bottom boundary conditions exchanged, though here we speak only in terms of internal heating.

The model studied is of interest for several reasons. First, convection that is wholly or partly driven by internal heating or cooling occurs in the Earth's mantle \citep{Schubert2001} and atmosphere \citep{Berlengiero2012}, other planetary atmospheres \citep{Ingersoll1978, Kaspi2009}, the cores of large main-sequence stars \citep{Kippenhahn1994}, and engineered systems involving exothermic chemical or nuclear reactions, including nuclear accident scenarios \citep{Asfia1996, Nourgaliev1997, Grotzbach1999}. Especially in the mantle and certain nuclear accidents, the upper boundary may be closer to a poor conductor than to the perfect conductor adopted in many models. Second, the convective configuration studied here is among the simplest possible in the sense that, when it is modelled using the Boussinesq equations, only two dimensionless parameters enter the dynamics (aside from any parameters used in describing the geometry). There are six configurations with this property \citep{Goluskin2015a} -- three instances of RB convection and three of IH convection -- and the present model is by far the least studied of the six. Finally, the model makes for an unusually tractable `textbook example' of fluid stability analysis; the linear and nonlinear stability thresholds are close but not identical, and both can be computed analytically for any boundary conditions on the velocity.

We are aware of only two studies of the present configuration \citep{Hewitt1980, Ishiwatari1994}. Both examined scale selection using two-dimensional simulations, and for free-slip boundaries \citet{Ishiwatari1994} used long-wavelength asymptotics to find the linear instability threshold of the static state and derive an asymptotic equation for the dynamics near onset. Beyond those studies, our results can be compared with work on RB convection between poorly conducting boundaries \citep[e.g.][]{Sparrow1963, Hurle1967, Otero2002, Johnston2009} and work on IH convection with a top that conducts perfectly, rather than poorly. The latter configuration was studied early on by \citet{Tritton1967}, \citet{Roberts1967}, and \citet{Thirlby1970}, in many subsequent works reviewed by \citet{Kulacki1985}, and more recently in simulations both at finite Prandtl numbers \citep{Ichikawa2006, CartlandGlover2009, Takahashi2010, Glover2013} and in the infinite limit \citep{Houseman1988, Schubert1993, Parmentier1994}.

Our mathematical model and its basic features are laid out in \S\ref{sec: model}. For various boundary conditions on the velocity, linear and nonlinear stability thresholds of the static state are found in \S\ref{sec: stab}. Integral quantities important to heat transport are addressed in \S\ref{sec: transport}, much of which is devoted to proving a lower bound on the mean temperature, and \S\ref{sec: conc} gives concluding remarks.

\section{The model}
\label{sec: model}

\begin{figure}
\begin{center}
\begin{tikzpicture}
\draw[thick] (-1.5,.5) -- (1.5,.5);
\draw[thick] (-1.5,-.5) -- (1.5,-.5);
\node at (0,.8) {$\partial_zT=-\Gamma$};
\node at (0,-.8) {$\partial_zT=0$};
\node at (0,0) {$H$};
\draw[latex-latex] (-1,-.5) -- (-1,.5);
\node at (-.8,0) {$d$};
\draw[-latex,thick] (.8,.25) -- (.8,-.25);
\node at (1.05,0) {$g$};
\end{tikzpicture}
\end{center}
\caption{Schematic of the convective configuration studied in the present work. Quantities shown are dimensional. The internal heat source ($H$) and gravitational acceleration ($g$) are constant and uniform.}
\label{fig: config}
\end{figure}
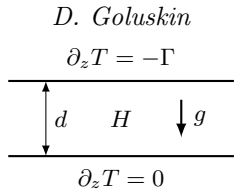

In dimensional terms, we are considering a fluid of thermal diffusivity $\kappa$ in a layer of height $d$, heated internally at rate $H$. The quantity $H$ has units of temperature per time and is equal to the specific rate of heating, normalized by density and specific heat. Figure \ref{fig: config} shows a schematic of this configuration. The perfectly insulating bottom boundary is enforced by a vanishing temperature flux, and the poorly conducting top boundary is modelled by a fixed heat flux, enforced by fixing the vertical temperature gradient to $-\Gamma$ there. The better the fluid transports heat, relative to the top boundary, the more accurate it is to model that boundary with a fixed heat flux \citep{Hurle1967}. Our model thus describes a situation where the fluid transports heat much better than the top boundary, which in turn transports heat much better than the bottom boundary. Any layer whose top boundary is much more conductive than its bottom one should be well described by our model whenever convection is sufficiently strong.

Statistically steady convection is possible only when the heat flux across the top boundary balances the internal heat production, hence we require
\begin{equation}
\kappa\Gamma=dH.
\end{equation}
The natural temperature scale in this system is
\begin{equation}
\Delta := d^2H/\kappa = d\Gamma. \label{eq: Delta}
\end{equation}
The quantity $d^2H/\kappa$ is the usual temperature scale of IH convection, and it agrees in this configuration with $d\Gamma$, the temperature scale of fixed-flux RB convection. Modelling the dynamics using the Boussinesq equations \citep{Spiegel1960, Chandrasekhar1981}, we nondimensionalize lengths by $d$, temperatures by $\Delta$, and times by the characteristic timescale of thermal diffusion, $d^2/\kappa$. The dimensionless dynamics are then governed by 
\begin{align}
\bnabla \bcdot \bu &= 0, \label{eq: inc} \\
\partial_t \bu + \bu \bcdot \bnabla \bu  &= 
	-\bnabla p + Pr \nabla^2 \bu + Pr R\,T \mathbf{\hat z}, \label{eq: u} \\
\partial_t T + \bu \bcdot \bnabla T& = \nabla^2 T + 1, \label{eq: T}
\end{align}
where $\mathbf u=(u,v,w)$ is the velocity of the fluid, $T$ is its temperature, and $p$ is its pressure. The heat source has been scaled to unit strength. The dimensionless control parameters, respectively called the Rayleigh and Prandtl numbers, are
\begin{align}
R :=& \frac{g\alpha d^3\Delta}{\kappa\nu}, & Pr :=& \frac{\nu}{\kappa}, \label{eq: R and Pr}
\end{align}
where $g$ is the constant gravitational acceleration acting in the $-\mathbf{\hat z}$ direction, $\alpha$ is the linear coefficient of thermal expansion, $\Delta$ is the temperature scale defined by (\ref{eq: Delta}), and $\nu$ is the kinematic viscosity. The above definition of $R$ agrees with the usual definitions of Rayleigh numbers as control parameters in IH convection \citep{Kulacki1985} and fixed-flux RB convection \citep{Sparrow1963}. As in RB convection, this $R$ is roughly the ratio of inertial forces to viscous forces and can be thought of as the strength with which the flow is driven.

The spatial domain of our model has a dimensionless vertical extent of $0\le z\le1$ and is infinite or periodic in its one or two horizontal directions. The dimensionless thermal boundary conditions are
\begin{equation}
\partial_zT|_{z=0} = 0, \qquad \partial_zT|_{z=1} = -1. \label{eq: T BC}
\end{equation}
For the velocity conditions at the top and bottom, we consider all four permutations of no-slip and free-slip boundaries, which are enforced by
\begin{align}
\text{no-slip:}\quad& u,\, v,\, w = 0, \label{eq: no-slip BC} \\
\text{free-slip:}\quad& \partial_zu,\, \partial_zv,\, w = 0. \label{eq: free-slip BC}
\end{align}

\begin{figure}
\begin{center}
(a)
\begin{tikzpicture}
\draw[white] (-1.5,-1.5) -- (-1.5,1.5);
\draw[gray,thick] (-1.5,.8) -- (1.5,.8);
\draw[gray,thick] (-1.5,-.8) -- (1.5,-.8);
\foreach \y in {-.8,-.79,...,.79}
\draw[black] (-1/1.6*\y*\y-\y+.4,\y) -- (-1/1.6*\y*\y-1/1.6*.02*\y-1/1.6*.0001-\y-.01+.4,\y+.01);
\draw[black,thick,-latex] (-1.3,-.8) -- (-1.3,-.2);
\draw[black,thick,-latex] (-1.3,-.8) -- (-.7,-.8);
\node at (0,1.1) {$\partial_zT=-1$};
\node at (0,-1.1) {$\partial_zT=0$};
\node at (-.8,-.55) {$T$};
\node at (-1.5,-.3) {$z$};
\end{tikzpicture}
\hspace{20pt}
(b)
\begin{tikzpicture}
\draw[white] (-1.5,-1.5) -- (-1.5,1.5);
\draw[gray,thick] (-1.5,.8) -- (1.5,.8);
\draw[gray,thick] (-1.5,-.8) -- (1.5,-.8);
\draw[black, rounded corners=2pt] (-.2,.8) -- (.2,.6) -- (.2,-.8);
\end{tikzpicture}
\end{center}
\caption{Schematics of mean vertical temperature profiles (a) in the static state and (b) as expected in strong convection. The dimensionless thermal boundary conditions are shown.}
\label{fig: profiles}
\end{figure}
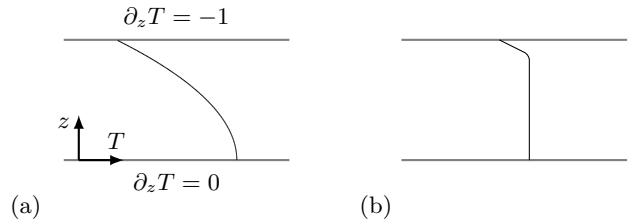

When the fluid is static, the unique steady temperature field is horizontally uniform and has the parabolic vertical profile
\begin{equation}
T_{st}(z) =\tfrac{1}{2}(1-z^2), \label{eq: T_st}
\end{equation}
as shown in figure \ref{fig: profiles}(a). By contrast, figure \ref{fig: profiles}(b) shows the sort of temperature profile that we expect at large $R$, where experience with similar systems suggests that strong convective mixing will render the fluid roughly isothermal outside of an upper thermal boundary layer. Integral quantities related to mean temperature profiles are discussed in~\S\ref{sec: transport}.

\begin{figure}
\begin{center}
\includegraphics[width=382pt]{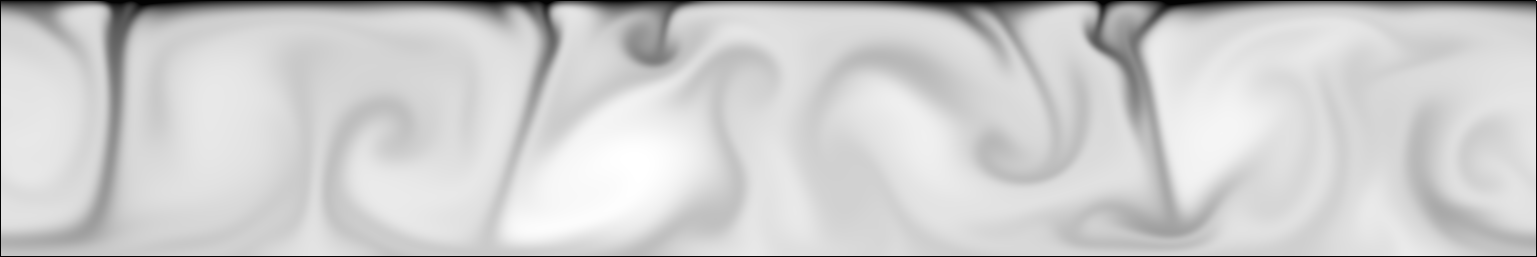}
\end{center}
\caption{Instantaneous temperature field from a two-dimensional simulation of our model with $R=1.44\cdot10^8$, $Pr=1$, a horizontal period of 6, and no-slip boundaries. The hottest fluid (white) is 0.06 dimensionless degrees warmer than the coldest fluid (black).}
\label{fig: field}
\end{figure}

To provide a concrete example of strong convection in our model, we carried out a two-dimensional simulation using {\tt nek5000} \citep{nek}. Figure \ref{fig: field} shows a typical temperature field from that simulation. As expected, cold plumes descend from an upper thermal boundary layer. We have not collected quantitative data on heat transport in this model, nor to our knowledge has anyone else.

\section{Stability of the static state}
\label{sec: stab}

To determine the stability of the static state, wherein $\bu=\mathbf0$ and $T=T_{st}$, we decompose the temperature field as $T(\bx,t)=T_{st}(z)+\theta(\bx,t)$, where $\theta$ is called the temperature fluctuation. Under the Boussinesq equations (\ref{eq: inc})-(\ref{eq: T}), fluctuations evolve according to
\begin{align}
\bnabla \bcdot \bu &= 0, \label{eq: inc pert} \\
\partial_t \bu + \bu \bcdot \bnabla \bu  &= 
	-\bnabla p + Pr \nabla^2 \bu + Pr R\, \theta \mathbf{\hat z}, \label{eq: u pert} \\
\partial_t\theta + \bu\bcdot\bnabla\theta &= \nabla^2\theta + zw, \label{eq: theta} 
\end{align}
where pressure has been redefined to absorbed the $T_{st}$ term. The stability of the static state is equivalent to the stability of the zero solution of the above fluctuation equations, and the latter is more convenient to analyse. Linear and nonlinear stability analyses will yield Rayleigh number thresholds for the static state -- $R_L$ and $R_E$, respectively -- such that $R<R_E$ suffices for global stability, and $R>R_L$ suffices for linear instability. Both the linear and nonlinear analyses lead to linear eigenproblems whose spectra must be determined. The derivations of these eigenproblems follow standard methods and are outlined in \S\ref{sec: linear eigen} and \S\ref{sec: energy eigen}. Both eigenproblems are solved exactly by asymptotic expansion in \S\ref{sec: asy}, which is possible here because the heat fluxes are fixed at both boundaries.

\subsection{Linear stability eigenproblem}
\label{sec: linear eigen}

To find a threshold for the linear instability of infinitesimal perturbations, we neglect the nonlinear terms in the fluctuation equations (\ref{eq: inc pert})-(\ref{eq: theta}). The first half of the procedure for finding $R_L$ closely follows the classic calculation for RB convection \citep{Rayleigh1916, Chandrasekhar1981}. The linearizations of (\ref{eq: theta}) and $\mathbf{\hat z}\cdot\nabla\times\nabla\times(\ref{eq: u pert})$ form a closed pair of evolution equations for $w$ and $\theta$ that have the same linear stability threshold as the full equations:
\begin{align}
\tfrac{1}{Pr}\partial_t\nabla^2w &= \nabla^4 w + R \nabla_H^2 \theta, \label{eq: w marginal}  \\
\partial_t\theta &= \nabla^2\theta + zw, \label{eq: theta marginal} 
\end{align}
where $\nabla_H^2:=\partial_x^2+\partial_y^2$ is the horizontal Laplacian operator. Regardless of whether the velocity boundary conditions are no-slip or free-slip,
\begin{align}
w,\, \partial_z\theta = 0 \text{ at } z=0,1.
\end{align}
The final two conditions on $w$ depend on whether the boundaries are no-slip or free-slip and are derived from the definitions (\ref{eq: no-slip BC})-(\ref{eq: free-slip BC}) with the help of the incompressibility condition (\ref{eq: inc}). We consider all four combinations here:
\begin{align}
\text{no-slip:}~~& w'|_{z=0},~\;w'|_{z=1}=0, \label{eq: no-slip} \\
\text{free-slip top:}~~& w'|_{z=0},~\,w''|_{z=1}=0, \label{eq: free-slip top}  \\
\text{free-slip bottom:}~~& w''|_{z=0},~\,w'|_{z=1}=0, \label{eq: free-slip bottom} \\
\text{free-slip:}~~& w''|_{z=0},~w''|_{z=1}=0, \label{eq: free-slip}
\end{align}
where primes denote $\partial_z$. To apply our results to the dynamically equivalent system of internally \emph{cooled} fluid with an insulating top and poorly conducting bottom, we need only remember that condition (\ref{eq: free-slip top}) would correspond to a free-slip bottom and condition (\ref{eq: free-slip bottom}) to a free-slip top.

The righthand side of (\ref{eq: w marginal})-(\ref{eq: theta marginal}) can be regarded as a linear operator acting on $[\nabla^2w~\,\theta]^T$. At the stability threshold we seek, the spectrum of the operator is marginally stable, meaning at least one eigenvalue has a vanishing real part. Here we look only for marginally stable states that are stationary, as opposed to time-dependent. Such time-independent states obey the linear eigenproblem
\begin{align}
\nabla^4 w &= -R \nabla_H^2 \theta, \label{eq: w stationary} \\
\nabla^2\theta &= -zw. \label{eq: theta stationary}
\end{align}
We define $R_L$ as the smallest $R$ at which a stationary, marginally stable state exists:
\begin{equation}
R_L := \inf\left\{ R ~\big|~ \text{(\ref{eq: w stationary})-(\ref{eq: theta stationary}) has a nonzero solution} \right\}. \label{eq: R_L def}
\end{equation}
The $R$ for which nonzero solutions exist are generalized eigenvalues; at such an $R$ there is a zero in the spectrum of an operator taking the form $\mathcal A-R\mathcal B$, where $\mathcal A$ and $\mathcal B$ are linear differential operators.

The condition $R>R_L$ is sufficient for linear instability, but, because we have assumed stationarity, it may not be necessary. Showing that it is necessary would require proving that all marginal states are indeed stationary. This is fairly easy in RB convection \citep{Pellew1940}, but the analogous proof fails here because of the non-constant coefficient in equation (\ref{eq: theta marginal}). A functional analytical approach has been used to prove stationarity in certain IH configurations \citep{Herron2001a, Herron2003} but apparently not for fixed-flux thermal boundary conditions.

Because the eigenproblem (\ref{eq: w stationary})-(\ref{eq: theta stationary}) is linear and lacks horizontal boundaries, we can Fourier transform it in $x$ and $y$ (or, equivalently, apply a normal mode substitution). This yields a separate eigenproblem for each magnitude, $k$, of the horizontal wavevector:
\begin{align}
\hat w^{(4)} -2k^2\hat w''+k^4\hat w &= R k^2 \hat\theta, \label{eq: w marginal k}  \\
\hat\theta''-k^2\hat\theta &= -z\hat w, \label{eq: theta marginal k} 
\end{align}
where $\hat w(z)$ and $\hat\theta(z)$ can be complex but obey the same boundary conditions as $w$ and $\theta$. These ordinary differential eigenproblems have discrete spectra, but the union of their spectra over all possible $k$ is the same as the spectrum of the original partial differential eigenproblem (\ref{eq: w stationary})-(\ref{eq: theta stationary}). Expression (\ref{eq: R_L def}) for $R_L$ thus becomes
\begin{equation}
R_L = \inf_{k^2>0}\min\left\{ R ~\big|~ \text{(\ref{eq: w marginal k})-(\ref{eq: theta marginal k}) has a nonzero solution} \right\}, \label{eq: R_L k}
\end{equation}
where $k^2$ cannot be zero because horizontally uniform $w$ would violate incompressibility.

In similar models of IH convection, the value of $R_L$ must be found by solving the eigenproblem (\ref{eq: w marginal k})-(\ref{eq: theta marginal k}) numerically for various fixed $k$ \citep{Roberts1967, Kulacki1975a}. In the present case, we have carried out such numerics only to confirm that, for all four pairs of velocity conditions, the generalized eigenvalue $R$ decreases monotonically as $k^2\to0$. The infimum of expression (\ref{eq: R_L k}) can thus be replaced by the long-wavelength limit,
\begin{equation}
R_L = \lim_{k^2\to0}\min\left\{ R ~\big|~ \text{(\ref{eq: w marginal k})-(\ref{eq: theta marginal k}) has a nonzero solution} \right\}, \label{eq: R_L lim}
\end{equation}
and an exact analytical expression for $R_L$ can be found by asymptotically expanding the eigenproblem in $k^2$. This has been done for free-slip boundaries by \citet{Ishiwatari1994} and is carried out for other velocity conditions in \S\ref{sec: asy}.

Monotonic decrease of the generalized eigenvalue $R$ as $k^2\to0$ has been found in various other convective systems where heat fluxes are fixed on both boundaries \citep{Sparrow1963, Chapman1980a, Depassier1982}. We do not know of any analytical proofs of this feature, though it seems to be a fairly general consequence of such boundary conditions.

\subsection{Energy stability eigenproblem}
\label{sec: energy eigen}

Returning to the nonlinear fluctuation equations (\ref{eq: inc pert})-(\ref{eq: theta}), we now seek a Rayleigh number, $R_E$, below which the static state is globally stable. As in most studies of fluid stability, we prove such a threshold using the energy method \citep{Serrin1959, Joseph1976, Straughan2004}. In particular, we follow \citet{Joseph1965} in considering (generalized) energies of the form
\begin{equation}
E_\gamma[\bu,\theta](t) : = \tfrac{1}{2}\fint\left( \tfrac{1}{Pr\,R}|\bu|^2
	+ \gamma\theta^2\right)d\bx, \label{eq: E_gamma}
\end{equation}
where $\gamma>0$ is a coupling parameter to be chosen later, and where $\fint$ denotes an average over the volume. The stability of the static state follows if the energy is a Lyapunov functional -- that is, if
\begin{align}
E_\gamma[\bu,\theta] &\ge 0, \label{eq: Lyap 1} \\
\tfrac{d}{dt}E_\gamma[\bu,\theta] &\le 0 \label{eq: Lyap 2}
\end{align}
for all possible $\bu$ and $\theta$, with equality holding only when both arguments are zero. The first condition always holds here since all parameters in definition (\ref{eq: E_gamma}) are positive. The second condition cannot hold when $R>R_L$ since nonlinear stability would be inconsistent with linear instability. The most we hope for is finding a threshold $R_{E_\gamma}$, where $R_{E_\gamma}\le R_L$, such that $R<R_{E_\gamma}$ is a sufficient condition for the second Lyapunov condition (\ref{eq: Lyap 2}) to hold.

We prove stability up to the largest threshold we can by choosing the value of $\gamma$ that maximizes $R_{E_\gamma}$. An even larger threshold might be proven by optimizing over a broader family of Lyapunov functions than the ansatz (\ref{eq: E_gamma}). However, the condition (\ref{eq: Lyap 2}) is generally very hard to check for a candidate functional. Like most authors, with a few exceptions \citep{Kaiser2005, Huang2015}, we have avoided this difficulty by the energy method, which entails restricting ourselves to Lyapunov functions that (1) are quadratic in the fluctuation variables, and (2) are conserved by the nonlinear terms of the evolution equations (\ref{eq: inc pert})-(\ref{eq: theta}). For such energies, a condition on the spectrum of a linear eigenproblem, introduced below, suffices to guarantee that $\tfrac{d}{dt}E_\gamma\le0$.

The spectral condition that is sufficient for Lyapunov stability has been derived and solved numerically in similar IH configurations \citep{Kulacki1975a, Straughan1990}. We can see how the eigenproblem arises by adding the volume averages of $\frac{1}{Pr\,R}\bu\cdot(\ref{eq: u pert})$ and $\gamma\theta\times(\ref{eq: theta})$, and then integrating by parts to find
\begin{equation}
\tfrac{d}{dt}E_\gamma = -\fint\left[
	\tfrac{1}{R}|\bnabla\bu|^2+\gamma|\bnabla\theta|^2-\left(1+\gamma z\right)w\theta\right]d\bx.
	\label{eq: dE/dt}
\end{equation}
Relaxing the dynamical constraints on $\bu$ and $\theta$ in the above expression gives
\begin{equation}
\tfrac{d}{dt}E_\gamma \le -\inf_{\substack{\bu,\theta\in H^2 \\ \nabla\cdot\bu=0 \\ \text{BCs}}}
	\left\{\fint\left[
	\tfrac{1}{R}|\bnabla\bu|^2+\gamma|\bnabla\theta|^2-\left(1+\gamma z\right)w\theta\right]d\bx\right\}, 
\end{equation}
where the infimum of the functional is over sufficiently smooth $\bu$ and $\theta$ that are subject only to incompressibility and the dynamical boundary conditions. The Euler-Lagrange equations of this functional, like the linear stability equations, can be reduced to a closed system for $w$ and $\theta$,
\begin{align}
\nabla^4w &= -\tfrac{1}{2}R(1+\gamma z)\nabla_H^2\theta, \label{eq: w EL} \\
\gamma\nabla^2\theta &= -\tfrac{1}{2}(1+\gamma z)w, \label{eq: theta EL}
\end{align}
where $w$ and $\theta$ in the Euler-Lagrange equations obey the same boundary conditions as the dynamical variables of the same name. If $R$ lies below the spectrum of this eigenproblem, then $\tfrac{d}{dt}E_\gamma\le0$. \citep[This is demonstrated by][for example, using a rescaling of $\theta$ that simplifies the argument.]{Straughan1990} The condition $R<R_{E_\gamma}$ thus suffices for Lyapunov stability, where
\begin{equation}
R_{E_\gamma} := \inf\left\{ R ~\big|~ \text{(\ref{eq: w EL})-(\ref{eq: theta EL}) has a nonzero solution} \right\}. \label{eq: R_E_gamma def}
\end{equation}

As in the linear stability analysis of \S\ref{sec: linear eigen}, we can Fourier transform in the horizontal directions to get an ordinary differential equation eigenproblem for each horizontal wavevector magnitude, $k$:
\begin{align}
\hat w^{(4)} -2k^2\hat w''+k^4\hat w &= \tfrac{1}{2}R k^2(1+\gamma z)\hat\theta,
	\label{eq: w EL k}  \\
\gamma(\hat\theta''-k^2\hat\theta) &= -\tfrac{1}{2}(1+\gamma z)\hat w.
	\label{eq: theta EL k} 
\end{align}
Expression (\ref{eq: R_E_gamma def}) then becomes
\begin{equation}
R_{E_\gamma} = \inf_{k^2>0}\min\left\{ R ~\big|~ \text{(\ref{eq: w EL k})-(\ref{eq: theta EL k}) has a nonzero solution} \right\}. \label{eq: R_E_gamma k}
\end{equation}
Since $E_\gamma$ is only certain to be a valid Lyapunov functional when $R<R_{E_\gamma}$, we get the strongest result by choosing $\gamma$ to maximize $R_{E_\gamma}$. This optimized threshold is what we call $R_E$. That is,
\begin{equation}
R_E := \max_{\gamma>0}\inf_{k^2>0}\min\left\{ R ~\big|~ \text{(\ref{eq: w EL k})-(\ref{eq: theta EL k}) has a nonzero solution} \right\}. \label{eq: R_E def}
\end{equation}
As in the linear stability analysis, the infimum in the above expression occurs as $k^2\to0$. We have confirmed this statement, at least for the optimal values of $\gamma$ that we eventually choose, by numerically solving the eigenproblem (\ref{eq: w EL k})-(\ref{eq: theta EL k}) for various $k$. Again we can replace the infimum with the $k^2\to0$ limit,
\begin{equation}
R_E = \max_{\gamma>0}\lim_{k^2\to0}\min\left\{ R ~\big|~ \text{(\ref{eq: w EL k})-(\ref{eq: theta EL k}) has a nonzero solution} \right\}, \label{eq: R_E}
\end{equation}
and we can solve the energy stability eigenproblem by expanding asymptotically in $k^2$.

\subsection{Analytical solution of the stability eigenproblems}
\label{sec: asy}

To evaluate expression (\ref{eq: R_L lim}) for $R_L$ and expression (\ref{eq: R_E}) for $R_E$, we expand the eigenproblems (\ref{eq: w marginal k})-(\ref{eq: theta marginal k}) and (\ref{eq: w EL k})-(\ref{eq: theta EL k}), respectively, in the small quantity $k^2$. Long-wavelength expansions have been applied previously to convective models with fixed-flux thermal boundary conditions, both to find $R_L$ and to capture the nonlinear dynamics near onset \citep{Childress2004, Chapman1980, Chapman1980a, Ishiwatari1994}, although we are not aware of their use in finding $R_E$.

Anticipating that $O(\hat w)=k^2O(\hat\theta)$ in the asymptotic solutions of both eigenproblems, we apply the expansions
\begin{align}
\hat w(z) &= k^2W_0(z) + k^4W_2(z) + \cdots, \label{eq: w expansion} \\
\hat \theta(z) &= \theta_0(z) + k^2\theta_2(z) + \cdots, \\
R &= R_0 + k^2R_2 + \cdots, \label{eq: R expansion}
\end{align}
where $R_0=R_L$ in the linear analysis, and $R_0=R_{E_\gamma}$ in the energy analysis. From the eigenproblems (\ref{eq: w marginal k})-(\ref{eq: theta marginal k}) and (\ref{eq: w EL k})-(\ref{eq: theta EL k}), we need the $\hat\theta$ equations only at $O(1)$ and $O(k^2)$ and the $\hat w$ equations only at $O(1)$:
\begin{align}
\begin{array}{r}
\text{linear\,} \\ \text{analysis:}
\end{array}
&& \theta_0'' &= 0, & ~~
W_0^{(4)} &= R_L\theta_0, & ~~
\theta_2'' &= \theta_0 - zW_0, \label{eq: R_L triplet} \\
\begin{array}{r}
\text{energy\,} \\ \text{analysis:}
\end{array}
&& \theta_0'' &= 0, & W_0^{(4)} &= \tfrac{1}{2}R_{E_\gamma}(1+\gamma z)\theta_0, 
	& \theta_2'' &= \theta_0 - \tfrac{1}{2}(1/\gamma+z)W_0, \label{eq: R_E triplet}
\end{align}
where all $W_n$ and $\theta_n$ satisfy the same boundary conditions as $w$ and $\theta$. 

In both the linear and energy analyses, the $\theta_0$ equations and their boundary conditions require that $\theta_0$ be constant. The nonzero constants arbitrarily fix the magnitudes of the eigenfunctions, so we take $\theta_0\equiv1$ for convenience. The $W_0$ equations give
\begin{equation}
W_0(z) = \begin{cases}
R_LP(z) & \text{linear analysis} \\
R_{E_\gamma}Q_\gamma(z) & \text{energy analysis},
\end{cases}
\end{equation}
where $P(z)$ and $Q_\gamma(z)$ are the unique polynomials of orders 4 and 5, respectively, that satisfy the $w$ boundary conditions and 
\begin{align}
P^{(4)}(z) &= 1, \\
Q_\gamma^{(4)}(z) &= \tfrac{1}{2}(1+\gamma z).
\end{align}
The appendix gives $P(z)$ and $Q_\gamma(z)$ for all four pairs of velocity conditions. Finally, the $\theta_2$ equations provide consistency conditions that can be solved for $R_L$ and $R_{E_\gamma}$. Since the fixed-flux boundary conditions require that $\int_0^1\theta_2''(z)dz$ vanish, the $\theta_2$ equations can be integrated and rearranged to find
\begin{align}
R_L &= \dfrac{1}{\int_0^1zP(z)dz}, \label{eq: R_L integral} \\
R_{E_\gamma} &= \dfrac{2}{\int_0^1\left(1/\gamma+z\right)Q_\gamma(z)dz}. \label{eq: R_E integral}
\end{align}

\begin{table}
\begin{center}
\begin{tabular}{rccc}
				& $R_L$& $R_E$ 	& \multicolumn{1}{c}{\quad~~gap} \\
\hline
no-slip			& 1440	& 1429.86	& 0.704\,\% \\
free-slip top		& 576	& 573.391	& 0.453\,\% \\
free-slip bottom	& 720	& 714.929	& 0.704\,\% \\
free-slip			& 240	& 239.055	& 0.394\,\%
\end{tabular}
\end{center}
\caption{Rayleigh numbers above which the static state is linearly unstable ($R_L$) and below which the static state is Lyapunov stable ($R_E$), along with the the percentage of $R_L$ by which $R_E$ falls short of $R_L$. Exact expressions for $R_E$ are given in table \ref{tab: R_E}. The finding $R_L=240$ for free-slip boundaries agrees with \citet{Ishiwatari1994}.}
\label{tab: stab}
\end{table}

\setlength\extrarowheight{8pt}
\begin{table}
\begin{center}
\begin{tabular}{rccc}
&	$R_{E_\gamma}$ & $\gamma^*$ & $R_E$  \\[2pt]
\hline
no-slip &
$\dfrac{100\,800\,\gamma}{9\gamma^2+35\gamma+35}$ &
$\frac{\sqrt{35}}{3}$ &
$2880\big(6\sqrt{35}-35\big)$ \\
free-slip top &
$\dfrac{403\,200\,\gamma}{99\gamma^2+350\gamma+315}$ &
$\frac{\sqrt{35}}{\sqrt{11}}$ &
$360\big(9\sqrt{385}-175\big)$ \\
free-slip bottom &
$\dfrac{403\,200\,\gamma}{64\gamma^2+280\gamma+315}$ &
$\frac{3\sqrt{35}}{8}$ &
$1440\big(6\sqrt{35}-35\big)$ \\
free-slip &
$\dfrac{30\,240\,\gamma}{16\gamma^2+63\gamma+63}$ &
$\frac{3\sqrt{7}}{4}$ &
$1440\big(8\sqrt{7}-21\big)$
\end{tabular}
\end{center}
\caption{The threshold ($R_{E_\gamma}$) below which the energy $E_\gamma$ is proven to be a Lyapunov functional for the static state, the optimal coupling parameter ($\gamma^*$) that maximizes this threshold, and the maximized threshold ($R_E$). Numerical approximations of $R_E$ are given in table \ref{tab: stab}.}
\label{tab: R_E}
\end{table}
\setlength\extrarowheight{0pt}

Values of $R_L$ for various boundary conditions on the velocity are given in table \ref{tab: stab}. These result from evaluating the integral (\ref{eq: R_L integral}) with the $P(z)$ given in the appendix. Since no-slip boundaries exert stresses that slow the fluid, it is unsurprising that the Rayleigh number needed to induce convection is smallest when the velocity conditions are both free-slip, larger when the conditions are mixed, and larger still when the conditions are both no-slip. When the velocity boundary conditions are mixed, $R_L$ is smaller when the top boundary is the free-slip one. This is reasonable because it is the unstable temperature gradient near the top boundary that drives the flow, and we expect from related studies of IH convection \citep{Kulacki1985} that mean velocities will be larger in the top half of the layer. A free-slip top thus encourages motion more than a free-slip bottom does.

Table \ref{tab: R_E} gives the expressions for $R_{E_\gamma}$ that are found by evaluating the integral (\ref{eq: R_E integral}) with the $Q_\gamma(z)$ given in the appendix. Also shown are the optimal coupling parameters, $\gamma^*$, that maximize these $R_{E_\gamma}$, as well as the maximized values, $R_E$.

The approximate numerical value of each $R_E$ is given alongside the corresponding $R_L$ in table \ref{tab: stab}. For each pair of velocity conditions, $R_E$ falls short of $R_L$ by less than 1\%. It is unknown whether subcritical convection can occur in the small gap between $R_E$ and $R_L$, though it can indeed occur when the top boundary is a perfect conductor, rather than a poor one \citep{Tveitereid1976, Busse2014}.

\section{Heat transport}
\label{sec: transport}

The stability analysis of the static state in \S\ref{sec: stab} determines whether convection occurs at a given $R$, except in the narrow range between $R_E$ and $R_L$. At $R$ large enough for convection to occur, we would like to predict quantitative features of the flow, especially quantities related to vertical heat transport.

Net heat transport is fixed in our model, being equal to both the flux at the top boundary and the rate of internal heating. The relative contributions of convection and conduction to the net transport, on the other hand, are dynamically determined. Several integral quantities are useful in characterizing these contributions. One such quantity, the mean fluid temperature, we bound from below in \S\ref{sec: bound}. Other useful quantities are discussed in \S\ref{sec: int quant}.

For our notation, we let an overline denote an average over the horizontal directions and infinite time, while angle brackets denote an average over volume and infinite time. Assuming periodicity on a horizontal domain of $[0,L_x]\times[0,L_y]$,
\begin{align}
\overline f(z) &:= 
	\liminf_{\tau\to\infty}\frac{1}{\tau}\frac{1}{L_xL_y}\int_0^\tau dt
	\int_0^{L_y}dy\int_0^{L_x}dx\,f(\bx,t), \\
\lan f\ran &:= 
	\liminf_{\tau\to\infty}\frac{1}{\tau}\frac{1}{L_xL_y}\int_0^\tau dt
	\int_0^1dz\int_0^{L_y}dy\int_0^{L_x}dx\,f(\bx,t).
\end{align}
If the domain is infinite, averages over $x$ and $y$ can instead be defined as limits. Defining time averages using $\liminf$, as opposed to $\limsup$, gives us a stronger result since the lower bound we find would be the same in either case.

\subsection{Lower bound on mean temperature}
\label{sec: bound}

We now prove a lower bound on $\T$ -- the mean temperature of the fluid, relative to the mean temperature of the top boundary, $\oT_T$. The comparison to $\oT_T$ is crucial; the volume average of $T$ alone cannot change from its initial value and so says nothing about the flow. The quantity $\T$ satisfies the uniform bounds
\begin{equation}
0 < \T \le \tfrac{1}{3}. \label{eq: T trivial}
\end{equation}
The lower bound follows from expression (\ref{eq: T PI}) below. To derive the upper bound, we integrate $z^2$ against the $T$ equation (\ref{eq: T}) to find $1/3-\T=\lan zwT\ran$. Incompressibility then gives $\lan zwT\ran=\lan zw\theta\ran$, multiplying $\theta$ against the temperature fluctuation equation (\ref{eq: theta}) gives $\lan zw\theta\ran=\lan|\nabla\theta|^2\ran\ge0$, and combining these relations gives $\T\le1/3$.

The upper bound on $\T$ is saturated only by the static state, in which heat is transported up and out of the layer by conduction alone. When $R$ is raised, and some of the net heat transport is taken over by convection, $\T$ must fall. Since $R$ is proportional to the dimensional heating rate, $H$, this decrease in $\T$ might seem counterintuitive until we recall that temperature has essentially been normalized by its value in the static state. If $\T$ decreases as $R$ is raised, this means only that the dimensional mean temperature, $\Delta\T$, grows sublinearly with $H$.

Experience with other convective systems strongly suggests that the flow will become ever more energetic and complicated as $R\to\infty$, and meanwhile $\T\to0$. The belief that $\T$ vanishes is analogous to the belief that the Nusselt number grows unboundedly in RB convection, and we are not aware of any method for proving such claims. What we can prove, beyond the uniform bounds (\ref{eq: T trivial}), is a lower bound on how quickly $\T$ decreases toward zero. The proof occupies the remainder of this subsection, but its result at leading order in $R$ is simply
\begin{equation}
\T \gtrsim 1.28\,R^{-1/3}. \label{eq: bound leading}
\end{equation}
In dimensional terms, the above bound says that the mean temperature, relative to that of the top boundary, grows with the heating rate no slower than $H^{2/3}$. We cannot judge the tightness of this bound since we are unaware of any experimental studies of $\T$.

\subsubsection{Background decomposition}

We derive a lower bound on $\T$ using the \emph{background method} \citep{Doering1992, Constantin1996}. The result is proven only for \emph{no-slip} boundaries, like the related bound that has been proven for fixed-flux RB convection \citep{Otero2002}. The background method entails decomposing the temperature field into a chosen background profile, $\tau(z)$, and remaining part, $\Theta(\bx,t)$:
\begin{equation}
T(\bx,t) = \tau(z) + \Theta(\bx,t). \label{eq: background decomp}
\end{equation}
The bound we obtain depends on our choice of $\tau$. This $\tau$ does not generally solve the governing equations, in which case $\Theta$ does not obey the fluctuation equations (\ref{eq: inc pert})-(\ref{eq: theta}).

The background profile $\tau$ must satisfy three conditions. First, it must be continuous. Second, it must obey the same fixed-flux boundary conditions (\ref{eq: T BC}) as $T$, so that $\Theta$ satisfies the homogenous conditions
\begin{equation}
\partial_z\Theta = 0 \text{ at } z=0,1. \label{eq: homog BC}
\end{equation}
As explained shortly, the boundary conditions on $\tau$ do not actually constrain our choice of background profile since they can be met by vanishingly thin boundary layers that do not affect the resulting bound. Finally, we must choose a $\tau$ for which we can show that a particular functional $\cQ$, introduced below, is non-negative.

We will bound $\T$ subject not to its full dynamical constraints but only to incompressibility, boundary conditions, and three integral relations found by taking $\lan T\times(\ref{eq: T})\ran$, $\lan\tau\times(\ref{eq: T})\ran$, and $\lan \bu\cdot(\ref{eq: u})\ran$. After integration by parts, these relations are
\begin{align}
\T &= \lan|\bnabla T|^2\ran, \label{eq: T PI} \\
\lan\tau'\Theta'\ran &= \lan\tau-\tau_T\ran - \lan\tau'^2\ran + \lan\tau'w\Theta\ran, \label{eq: tau PI} \\
R\wT &= \lan|\bnabla\bu|^2\ran, \label{eq: u PI}
\end{align}
where primes denote ordinary or partial $z$-derivatives. Time derivatives have vanished from (\ref{eq: T PI})-(\ref{eq: u PI}) because the volume integrals of $|\bu|$ and $|T|$ are bounded uniformly in time, a fact that follows from the present analysis \citep[cf.][]{Doering1992}. Relations (\ref{eq: T PI}) and (\ref{eq: u PI}) are the IH convection analogs of the power integrals of RB convection \citep{Malkus1954, Howard1963}.

The quantity we seek to bound appears in relation (\ref{eq: T PI}), which can be expanded as
\begin{equation}
\T = \lan\tau'^2\ran + \lan|\bnabla\Theta|^2\ran + 2\lan\tau'\Theta'\ran. \label{eq: T expand}
\end{equation}
As done by \citet{Lu2004}, we apply relations (\ref{eq: tau PI})-(\ref{eq: u PI}) to (\ref{eq: T expand}) and find
\begin{equation}
\T = 2\lan\tau-\tau_T\ran - \lan\tau'^2\ran + \cQ, \label{eq: T prebound}
\end{equation}
where
\begin{equation}
\cQ := \tfrac{a}{R}\lan|\bnabla\bu|^2\ran + \lan|\bnabla\Theta|^2\ran + \lan(2\tau'-a)w\Theta\ran,
	\label{eq: Q}
\end{equation}
and $a>0$ is to be fixed later. We must choose an admissible $\tau$ for which we can show that $\cQ\ge0$, where $\cQ$ is treated as a functional of any $\Theta$ and incompressible $\bu$ that are sufficiently smooth and satisfy the dynamical boundary conditions. When $\cQ\ge0$, expression (\ref{eq: T prebound}) gives the bound
\begin{equation}
\lan T-\oT_T\ran \ge 2\lan\tau-\tau_T\ran - \lan\tau'^2\ran. \label{eq: T bound}
\end{equation}
Having already relaxed the full dynamical constraints, we further limit our scope to $\tau$ profiles that are piecewise linear. These simplifications lead to a suboptimal bound but let us reach it analytically, which is necessary for the bound to apply at arbitrarily \mbox{large $R$}.

\subsubsection{Piecewise linear background profile}
\label{sec: profile}

\begin{figure}
\begin{center}
\includegraphics[width=170pt]{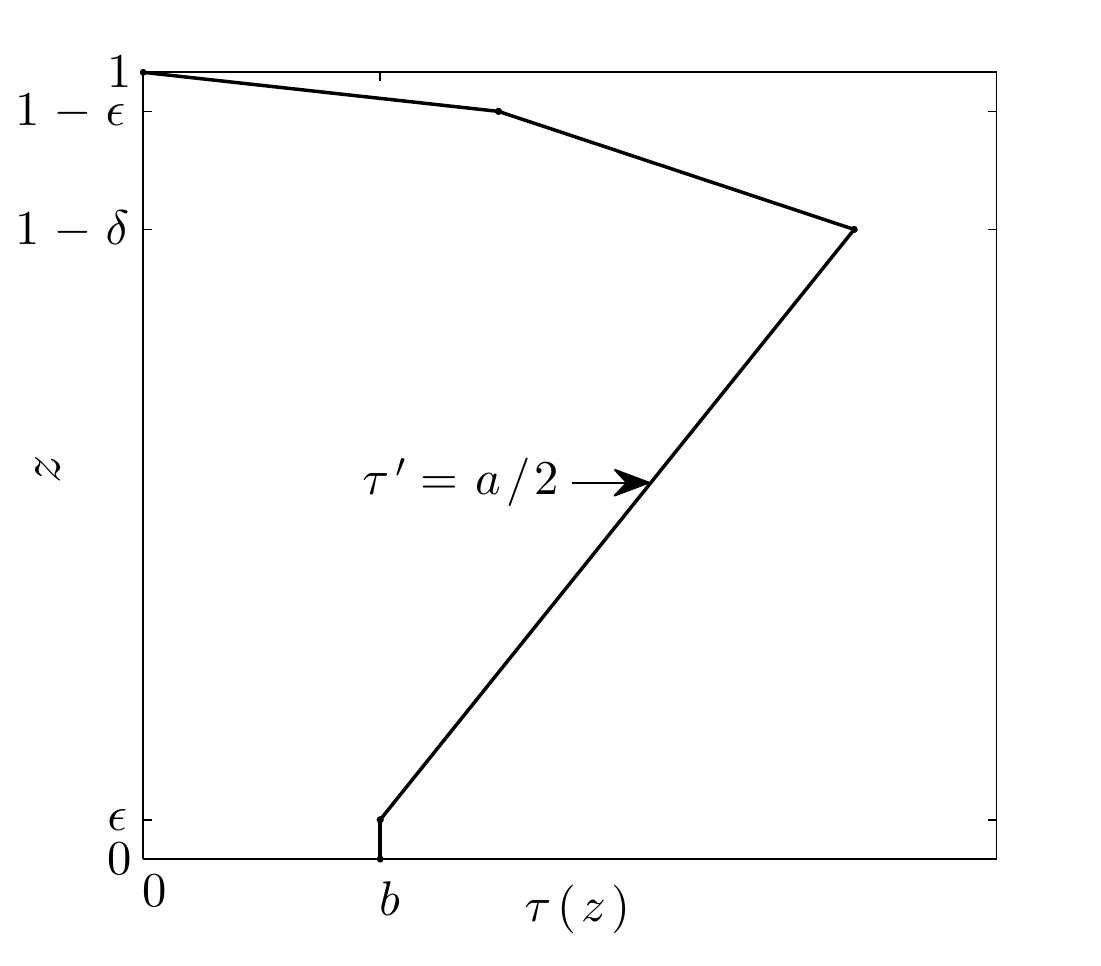}
\end{center}
\caption{Schematic of the background profile, $\tau(z)$, that we consider. The parameters $\delta$, $a$, and $b$ are optimized to maximize the lower bound (\ref{eq: T bound}) while maintaining the non-negativity of $\cQ$. We can neglect the two layers of thickness $\epsilon$ in our analysis (see text).}
\label{fig: tau schematic}
\end{figure}

Figure \ref{fig: tau schematic} shows the family of $\tau$ that we consider here. In principle, the boundary layers of thickness $\epsilon$ are needed to satisfy the fixed-flux thermal boundary conditions, so that $\Theta$ obeys the corresponding homogenous conditions. In practice, however, carrying $\epsilon$ through the analysis and then taking $\epsilon\to0$ yields the same bound as setting $\epsilon\equiv0$ at the start, so we simply do the latter. Our calculation thus excludes the $O(\epsilon)$ terms that would make it fully rigorous but arrives at the same result. The fact that the thermal boundary conditions on $\tau$ effectively can be ignored relies on the conditions being fixed-flux. When a boundary layer is used to meet a fixed-\emph{temperature} condition, its effect on the resulting bound does not generally vanish as its thickness goes to zero because its slope approaches infinity.

With $\epsilon\equiv0$, our ansatz for $\tau$ consists of only two linear pieces:
\begin{equation}
\tau(z) = \begin{cases}
\left[\tfrac{b}{\delta}+\tfrac{a}{2}\left(\tfrac{1}{\delta}-1\right)\right](1-z) & 
	1-\delta\le z\le1 \\
b+\tfrac{a}{2}z & 0\le z\le1-\delta, \label{eq: linear tau}
\end{cases}
\end{equation}
where figure \ref{fig: tau schematic} shows the geometric meanings of $\delta$, $a$, and $b$. The top temperature is fixed as $\tau_T=0$ for convenience since adding a constant to $\tau$ does not affect the bound (\ref{eq: T bound}). The upper piece of $\tau$ turns out to be a boundary layer because we must choose an expression for its thickness, $\delta$, that vanishes as $R\to\infty$. The lower piece of $\tau$ is chosen to have a slope of $a/2$, whatever the value of $a$ we fix later -- a known trick for making the sign-indefinite term of $\cQ$ vanish outside the boundary layer \citep{Constantin1996, Lu2004}. With the ansatz (\ref{eq: linear tau}) chosen for $\tau$, the lower bound (\ref{eq: T bound}) becomes
\begin{equation}
\lan T-\oT_T\ran \ge 
	b(2-\delta) + \tfrac{a}{2}(1-\delta) - 
	\left( \tfrac{a^2}{4}+ab \right)\left(\tfrac{1}{\delta}-1\right) - \tfrac{b^2}{\delta}.
	\label{eq: T bound 2}
\end{equation}

\subsubsection{Optimal parameter choices}

We seek the optimal parameters -- $\delta^*$, $a^*$, and $b^*$ -- that maximize the lower bound (\ref{eq: T bound}) while still letting us show $\cQ\ge0$. There is one optimality condition that is unaffected by the requirement that $\cQ\ge0$; the lower bound (\ref{eq: T bound 2}) is maximized when its partial derivative with respect to $a$ vanishes, and this requires that
\begin{equation}
b^*=\tfrac{1}{2}(\delta-a). \label{eq: b opt}
\end{equation}
Making this optimal choice, we eliminate $b$ from the bound to find
\begin{equation}
\lan T-\oT_T\ran \ge
	\tfrac{3}{4}\delta - \tfrac{1}{2}a-\tfrac{1}{2}\delta^2+\tfrac{1}{2}a\delta-\tfrac{1}{4}a^2.
\label{eq: T bound 3}
\end{equation}
It remains to choose $\delta$ and $a$ optimally, but first we must find conditions on these parameters that ensure $\cQ\ge0$.

Determining conditions sufficient for $\cQ\ge0$ requires some functional analysis to bound the magnitude of the sign-indefinite term, $\lan(2\tau'-a)w\Theta\ran$. Following \citet{Otero2002}, we can proceed in spectral space, noting that $\cQ$ is bounded below by an integral over horizontal wavevectors:
\begin{equation}
\cQ \ge \int_\mathbf k\cQk d\mathbf k,
\end{equation}
where
\begin{equation}
\cQk :=  \tfrac{a}{R}\lan\tfrac{1}{k^2}|\hat w_\bk''|^2 + 
		 2|\hat w_\bk'|^2 + k^2|\hat w_\bk|^2\ran +
		 \lan|\hat\Theta_\bk'|^2 + k^2|\hat\Theta_\bk|^2\ran +
		 \mathfrak R\lan (2\tau'-a) \widetilde w_\bk \hat\Theta_\bk\ran,
\end{equation}
and where $\hat w_\bk(z)$ and $\hat\Theta_\bk(z)$ are the horizontal Fourier transforms of $w$ and $\Theta$, $\widetilde w_\bk$ is the complex conjugate of $\hat w_\bk$, and $\mathfrak R$ denotes the real part of a complex quantity. Incompressibility has been used to eliminate horizontal velocity components from $\cQk$. The sign-indefinite term of $\cQk$ is nonzero only in the boundary layer, and its magnitude there is bounded by \citep{Otero2002}
\begin{equation}
\left|  \mathfrak R \lan (2\tau'-a) \widetilde w_\bk\hat\Theta_\bk\ran \right| 
	\le   \tfrac{\delta^2}{4\sqrt{2}}
		\left( \tfrac{\alpha}{k^2}\lan|\hat w_\bk''|^2\ran + 
		\beta\lan|\hat w_\bk'|^2\ran + \tfrac{1}{\beta}\lan|\hat\Theta_\bk'|^2\ran +
		\tfrac{k^2}{\alpha}\lan|\hat\Theta_\bk|^2\ran \right),
\end{equation}
where we have made use of the fact that $2\tau'-a=-1$ in the boundary layer for the optimal choice $b^*=\tfrac{1}{2}(\delta-a)$. The above estimate, and no other part of our proof, relies on the assumption of no-slip boundaries. The two previous expressions give
\begin{multline}
\cQk \ge \tfrac{1}{k^2}\left(\tfrac{a}{R}-\tfrac{\alpha\delta^2}{4\sqrt{2}}\right)
	\lan|\hat w_\bk''|^2\ran + 
	\left(\tfrac{2a}{R}-\tfrac{\beta\delta^2}{4\sqrt{2}}\right)\lan|\hat w_\bk'|^2\ran + \\
	\left(1-\tfrac{\delta^2}{4\sqrt{2}\beta}\right)\lan|\hat\Theta_\bk'|^2\ran +
	k^2\left(1-\tfrac{\delta^2}{4\sqrt{2}\alpha}\right)\lan|\hat\Theta_\bk|^2\ran.
\end{multline}
The non-negativity of all four coefficients in the above inequality suffices for the non-negativity of each $\cQk$ and, in turn, of $\cQ$. We choose $\alpha=\beta=\delta^2/4\sqrt{2}$, which is as large as the latter two coefficients allow. The non-negativity of the first coefficient, which also implies that of the second, then requires $a\ge\tfrac{1}{32}R\delta^4$. Our lower bound will be maximized by letting $\delta$ be as large as possible while guaranteeing $\cQ\ge0$, so we choose
\begin{equation}
a^*=\tfrac{1}{32}R\delta^4, \label{eq: a opt}
\end{equation}
after which the bound (\ref{eq: T bound 3}) becomes
\begin{equation}
\lan T-\oT_T\ran \ge
	\tfrac{3}{4}\delta - \tfrac{1}{64}R\delta^4 -\tfrac{1}{2}\delta^2 + 
	\tfrac{1}{64}R\delta^5 - \tfrac{1}{4096}R^2\delta^8.
\label{eq: T bound 4}
\end{equation}
To avoid both positive terms in the lower bound being subdominant when $R\to\infty$, we must choose $\delta$ no larger than $O(R^{-1/3})$. For such $\delta$, the last three terms are subdominant, so
\begin{equation}
\lan T-\oT_T\ran \gtrsim
	\tfrac{3}{4}\delta - \tfrac{1}{64}R\delta^4
\label{eq: T bound 5}
\end{equation}
at large $R$. The optimal $\delta^*$ that maximizes this leading-order expression is proportional to $R^{-1/3}$. Finding this $\delta^*$ and using it to put expressions (\ref{eq: a opt}) and (\ref{eq: b opt}) for $a^*$ and $b^*$ in terms of $R$ gives
\begin{align}
\delta^* &= 12^{1/3}R^{-1/3}, &
a^* &= \tfrac{3\cdot12^{1/3}}{8}R^{-1/3}, &
b^* &= \tfrac{5\cdot12^{1/3}}{16}R^{-1/3}.
\end{align}
With these parameter choices, $\tau'=-\tfrac{1}{2}+O(R^{-1/3})$ in the boundary layer. This is roughly half of what $\partial T/\partial z$ would be in the thermal boundary layer of an actual flow. In fixed-flux RB convection, on the other hand, the boundary layers of a similarly optimized $\tau$ profile have the \emph{same} $z$-derivative as the dynamical $T$ field at the boundaries.

Applying the optimal $\delta^*$ to the exact expression (\ref{eq: T bound 4}), we at last obtain our lower bound on the mean temperature,
\begin{equation}
\lan T-\oT_T\ran \ge \tfrac{9}{8}\left(\tfrac{3}{2}\right)^{1/3}R^{-1/3} - 
	\tfrac{89}{64}\left(\tfrac{3}{2}\right)^{2/3}R^{-2/3}.
\end{equation}
At large $R$, this bound scales like $R^{-1/3}$ and reduces to expression (\ref{eq: bound leading}). The lower bound on mean temperature proven by \citet{Lu2004} for a different internally heated configuration also scales like $R^{-1/3}$.

\subsection{Other quantities important to heat transport}
\label{sec: int quant}

We turn now to other integral quantities that, like $\T$, bear on the relative contributions of conduction and convection. The vertical heat flux, $J$, at a point is the sum of the fluxes due to conduction, $J_{cond}$, and convection, $J_{conv}$. In our nondimensionalization,
\begin{align}
J &= J_{cond}+J_{conv} \\
	&= -\partial_zT + wT.
\end{align}
The components of mean heat flux across a horizontal surface, $-\oT'(z)$ and $\owT(z)$, are not known \emph{a priori}, but their sum is; integrating the temperature equation (\ref{eq: T}) over the horizontal directions, time, and $[0,z]$ gives
\begin{equation}
\oJ(z)=-\oT'(z)+\owT(z)=z. \label{eq: oJ}
\end{equation}
This balance expresses the fact that, because the bottom boundary is insulating, the mean upward flux at height $z$ is equal to the rate, also $z$, at which heat is produced below that height. Integrating expression (\ref{eq: oJ}) over the vertical extent gives another useful balance,
\begin{equation}
\lan J\ran = \DT + \wT = \tfrac{1}{2}, \label{eq: mean J}
\end{equation}
where $\DT:=\oT_B-\oT_T$ is the difference between the mean temperatures at the bottom and top boundaries. That is, the mean heat flux over the layer is $1/2$ and is the sum of the conductive and convective parts, $\DT$ and $\wT$

\subsubsection{Mean temperature difference}
\label{sec: DT}

The relative contributions of conduction and convection can be characterized in similar but not identical ways by two quantities: the relative mean temperature of the fluid, $\T$, and the mean temperature difference between the boundaries, $\DT$, which is also the mean conductive transport across the layer. (We could equally well speak of $\wT$ instead of $\DT$ since we know they sum to $1/2$.) For $\DT$ we have the uniform upper bound
\begin{equation}
\DT\le\tfrac{1}{2}.
\end{equation}
A lower bound of zero seems likely, but we have not proven it. The upper bound follows from expressions (\ref{eq: u PI}) and (\ref{eq: mean J}). Much like $\T$, the quantity $\DT$ saturates its upper bound of $1/2$ only in the static state, and we expect but do not know how to prove that $\DT\to0$ as $R\to\infty$. In this limit, the net heat transport would be accomplished solely by convection, rather than conduction.

At large $R$, we expect $\DT$ and $\T$ to be even more similar. Strong convection typically renders $\oT(z)$ profiles roughly isothermal outside of boundary layers, as in the schematic of figure \ref{fig: profiles}(b), and here this would mean $\DT\sim\T$. Even with boundary conditions for which large-scale shear might keep the interior far from isothermal \citep{Goluskin2014, VanderPoel2014a}, we still expect $\DT$ and $\T$ to scale similarly at large $R$. Thus, since we have proven in \S\ref{sec: bound} that $\T$ can decay no faster than $R^{-1/3}$, it seems likely that the same is true of $\DT$.

Proving a parameter-dependent lower bound on $\DT$ remains an open challenge. The challenge is novel because $\DT$ is related to $\lan|\nabla \bu|^2\ran$, via expressions (\ref{eq: u PI}) and (\ref{eq: mean J}), but not to $\lan|\nabla T|^2\ran$. The background method would thus require decomposing $\bu$, whereas past applications of the method to convection have decomposed $T$ and bounded quantities related to $\lan|\nabla T|^2\ran$.

\subsubsection{Nusselt numbers and diagnostic Rayleigh numbers}
\label{sec: N}

In convective systems, the relative contributions of conduction and convection to net heat transport are often expressed using dimensionless \emph{Nusselt numbers}, $N$. One particular definition of $N$, together with a \emph{diagnostic Rayleigh number}, $Ra$, works well to reveal the parallels between RB configurations with different thermal boundary conditions \citep{Otero2002, Verzicco2008, Johnston2009, Wittenberg2010}. We define $N$ and $Ra$ in a way that agrees with these RB studies and extends to IH convection:
\begin{align}
N &:= \frac{\lan J\ran}{\lan J_{cond}\ran}, & 
Ra &:= R\frac{\lan J_{cond}\ran}{\lan J_{cond}\ran_{st}}, \label{eq: N Ra def}
\end{align}
where $\lan J\ran$ and $\lan J_{cond}\ran$ refer to the developed flow, while $\lan J_{cond}\ran_{st}$ refers to the static state. In our present model,
\begin{align}
N &= \frac{1}{2\DT} = \frac{1}{1-2\wT}, &
Ra &= R/N. \label{eq: N Ra}
\end{align}
The definition (\ref{eq: R and Pr}a) of the control parameter $R$ uses the dimensional temperature scale $\Delta$, given in expression (\ref{eq: Delta}), that is proportional to the temperature difference between the boundaries in the static state. The diagnostic parameter $Ra$ essentially replaces this static temperature difference with that in the developed flow. The two parameters agree in the static state, but $Ra<R$ in sustained convection.

Restated in terms of the $N$ we have defined, the basic features of $\DT$ described in \S\ref{sec: DT} are: $N=1$ in the static state, $N>1$ in sustained flows, and we expect $N$ to grow without bound as $R\to\infty$. These same statements apply to the usual Nusselt number of RB convection. We have further argued (without proof) in \S\ref{sec: DT} that $\DT$ can decay no faster than $R^{-1/3}$. This would be equivalent to $N$ growing no faster than $Ra^{1/2}$.

Since we have proven an $R$-dependent bound on $\T$, it is natural to ask whether the Nusselt number and diagnostic Rayleigh number could instead be generalized to IH convection in a way that invokes $\T$, rather than $\DT$. This leads us to define quantities that are like $N$ and $Ra$ but with averages weighted proportionally to height,
\begin{align}
\widetilde N &:= \frac{\lan zJ\ran}{\lan zJ_{cond}\ran}, &
\widetilde{Ra} &:= R\frac{\lan zJ_{cond}\ran}{\lan zJ_{cond}\ran_{st}}. \label{eq: N Ra tilde def}
\end{align}
Because $\oJ(z)=z$ here, weighting by height is equivalent to weighting by the mean vertical heat flux at each height. For our present boundary conditions,
\begin{align}
\widetilde N &= \frac{1}{3\T}, &
\widetilde{Ra} &= R/\widetilde N. \label{eq: N Ra tilde}
\end{align}
Expressed in these terms, our lower bound (\ref{eq: bound leading}) on $\T$ becomes an upper bound on $\widetilde{N}$,
\begin{equation}
\widetilde{N} \lesssim 0.132\,\widetilde{Ra}^{1/2}. 
\end{equation}
The above bound has the same scaling as bounds that have been proven in a variety of RB configurations, so long as the RB bounds are expressed using definition (\ref{eq: N Ra def}) for $N$ and $Ra$ \citep{Constantin1996, Plasting2003, Otero2002, Wittenberg2010}.

\section{Conclusions}
\label{sec: conc}

This work has addressed internally heated convection beneath a poor conductor, a basic but largely overlooked configuration. We have found differing linear and nonlinear stability thresholds for the static state. Simple exact expressions exist for both thresholds, which is rare in studies of fluid stability, and they have been found here using long-wavelength asymptotics. Beyond the static state, we have bounded the mean temperature of the convecting fluid, relative to that of the top boundary, assuming no-slip velocity conditions on both boundaries. As the heating rate ($H$) is raised, this dimensional mean temperature can grow no slower than $H^{2/3}$. In terms of a dimensionless Nusselt number ($\widetilde N$) and diagnostic Rayleigh number ($\widetilde{Ra}$) that we have defined using the mean temperature, our bound takes the same form as upper bounds on Nusselt numbers in various other convective models.

Many fundamental features of the model studied here are yet to be explored. At Rayleigh numbers between our thresholds for global stability and linear instability, it is not known whether subcritical convection can be sustained, although the answer is affirmative when the upper boundary conducts perfectly instead of poorly \citep{Tveitereid1976}. While we have proved a parameter-dependent lower bound on the mean temperature of sustained flow, only uniform bounds are known for the mean temperature difference between the top and bottom boundaries. The latter quantity is useful in characterizing heat transport in the fluid, and it is much easier to measure experimentally than the mean fluid temperature. A bound on this temperature difference would amount to a bound on viscous dissipation, rather than on thermal dissipation, and a method for constructing it would likely yield novel results in other internally heated configurations as well \citep{Goluskin2015a}. Beyond analytical results, it seems the only studies of our configuration have been two-dimensional simulations at small-to-moderate Rayleigh numbers \citep{Hewitt1980, Ishiwatari1994}. The parameter regimes in which convection is strong and complicated are wide open for numerical simulations and laboratory experiments.

\begin{acknowledgments}
The author thanks Charles Doering and the anonymous referees for some very helpful comments on the manuscript. The author was supported during part of this research by the US National Science Foundation (NSF) Mathematical Physics award PHY-1205219.
\end{acknowledgments}

\appendix

\section{Polynomial eigenfunctions}
\label{app: poly}

The asymptotic solution of the linear stability eigenproblem in \S\ref{sec: asy} makes use of the unique fourth-order polynomials that satisfy $P^{(4)}(z)=1$ and the velocity boundary conditions (\ref{eq: no-slip})-(\ref{eq: free-slip}). These  polynomials are
\begin{equation}
P(z) = \begin{cases}
\frac{1}{24}\left(z^4-2 z^3+z^2\right) & \text{no-slip} \vspace{2pt} \\
\frac{1}{24}\left(z^4-\frac{5}{2}z^3+\frac{3}{2}z^2\right) & \text{free-slip top} \vspace{2pt} \\
\frac{1}{24}\left(z^4-\frac{3}{2}z^3+\frac{1}{2}z\right) & \text{free-slip bottom} \vspace{2pt} \\
\frac{1}{24}\left(z^4-2 z^3+z\right) & \text{free-slip}.
\end{cases}
\end{equation}
The same $P(z)$ are given by \citet{Chapman1980}, scaled for their domain of $-1\le z\le1$ rather than our domain of $0\le z\le1$. The asymptotic solution of the energy stability eigenproblem requires the unique fifth-order polynomials that satisfy $Q_\gamma^{(4)}(z)=\tfrac{1}{2}(1+\gamma z)$ and the velocity boundary conditions. These polynomials are

\begin{equation}
Q_\gamma(z) = \begin{cases}
\frac{1}{240}\left[\gamma  z^5+5z^4-(3\gamma+10)z^3+(2\gamma+5)z^2\right]
	& \text{no-slip} \vspace{2pt} \\
\frac{1}{240}\left[\gamma  z^5+5z^4-\left(\frac{9}{2}\gamma+\frac{25}{2}\right)z^3+
	\left(\frac{7}{2}\gamma+\frac{15}{2}\right)z^2\right]
	& \text{free-slip top} \vspace{2pt} \\
\frac{1}{240}\left[\gamma z^5+5z^4-\left(2\gamma+\frac{15}{2}\right)z^3+
	\left(\gamma+\frac{5}{2}\right)z\right] 
	& \text{free-slip bottom} \vspace{2pt} \\
\frac{1}{240}\left[\gamma z^5+5z^4-\left(\frac{10}{3}\gamma+10\right)z^3
	+\left(\frac{7}{3}\gamma+5\right)z\right]
	& \text{free-slip}.
\end{cases}
\end{equation}

\begin{figure}
(a)\includegraphics[width=185pt]{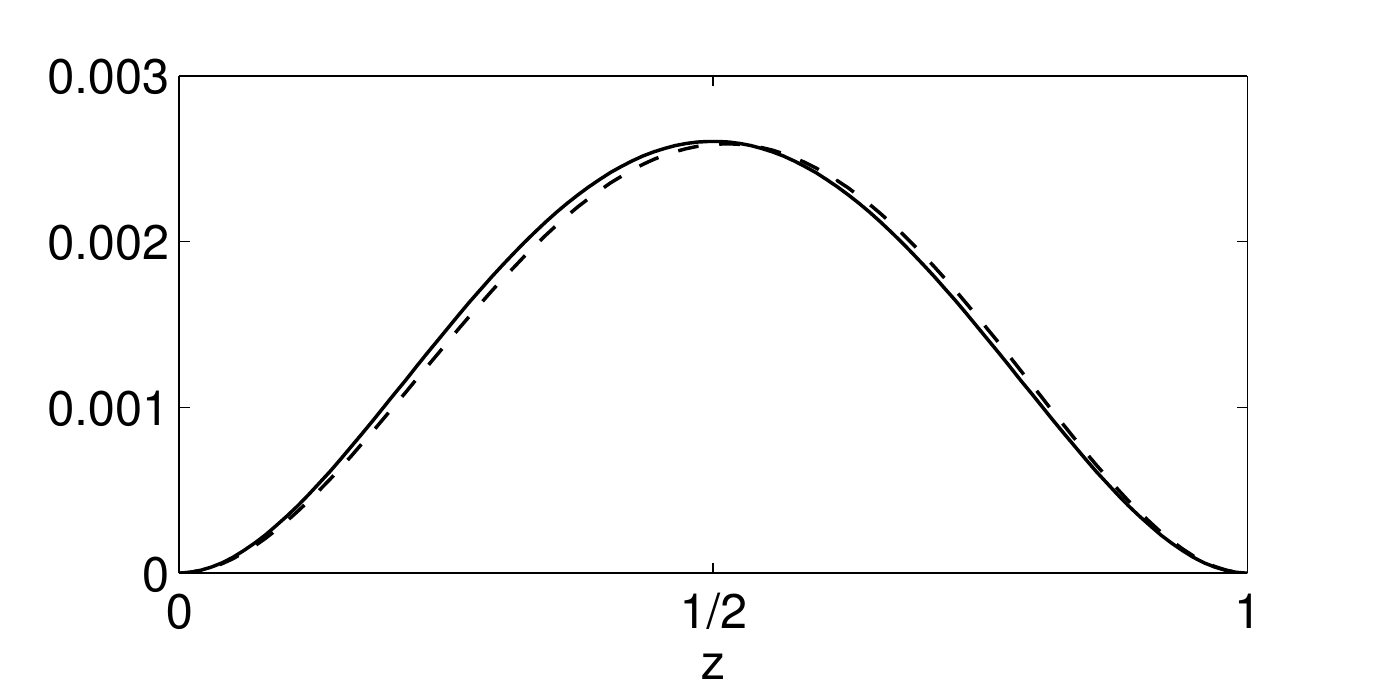}
(b)\includegraphics[width=185pt]{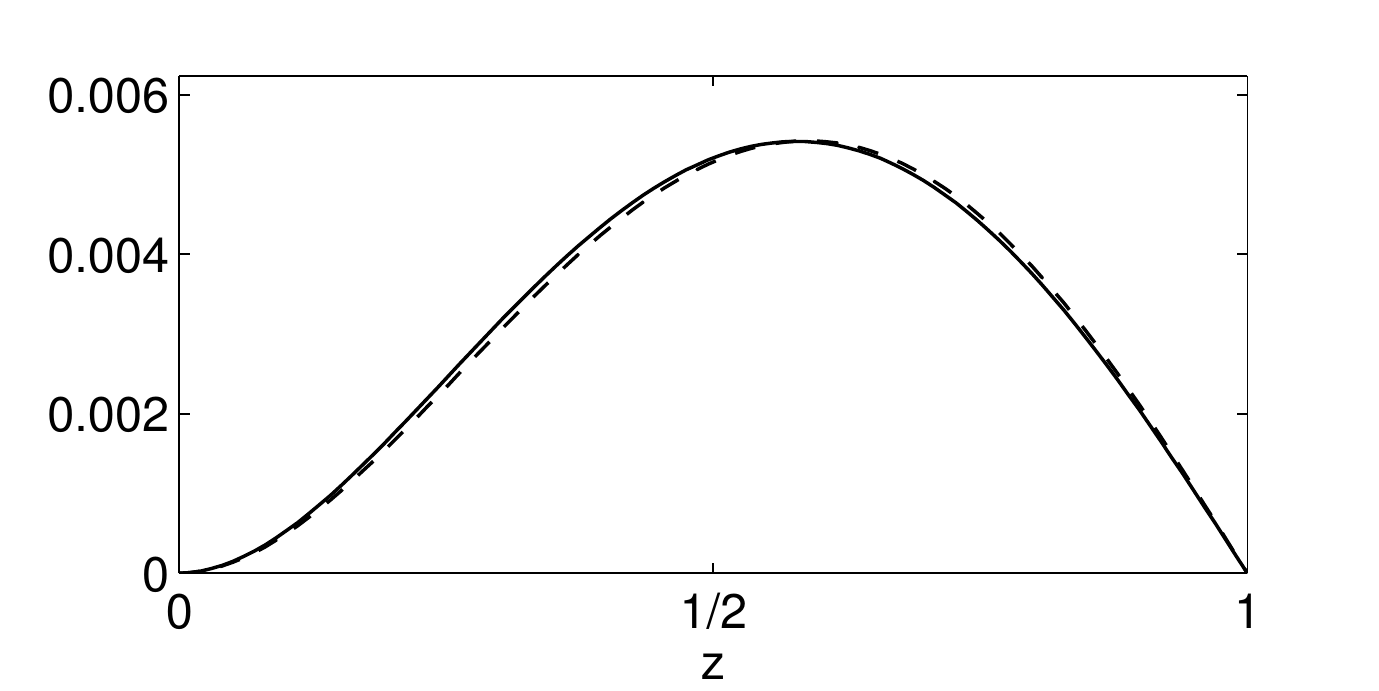} \\
(c)\includegraphics[width=185pt]{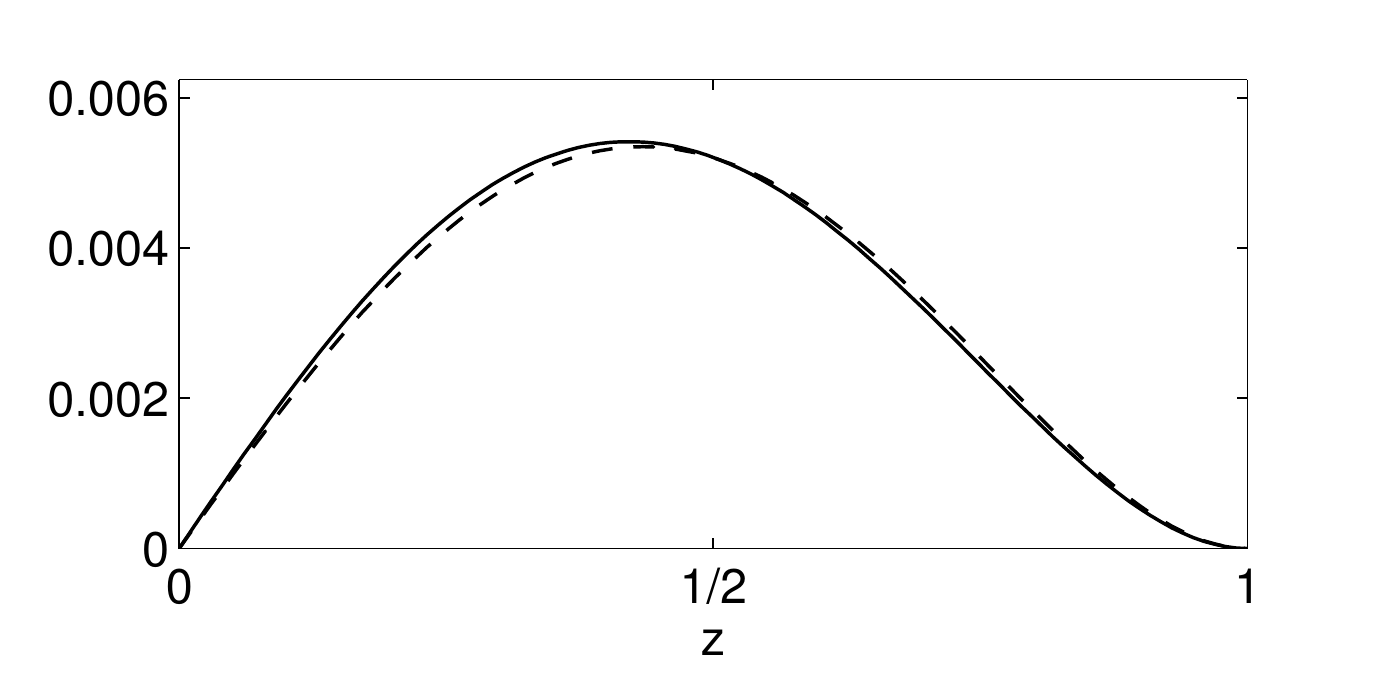}
(d)\includegraphics[width=185pt]{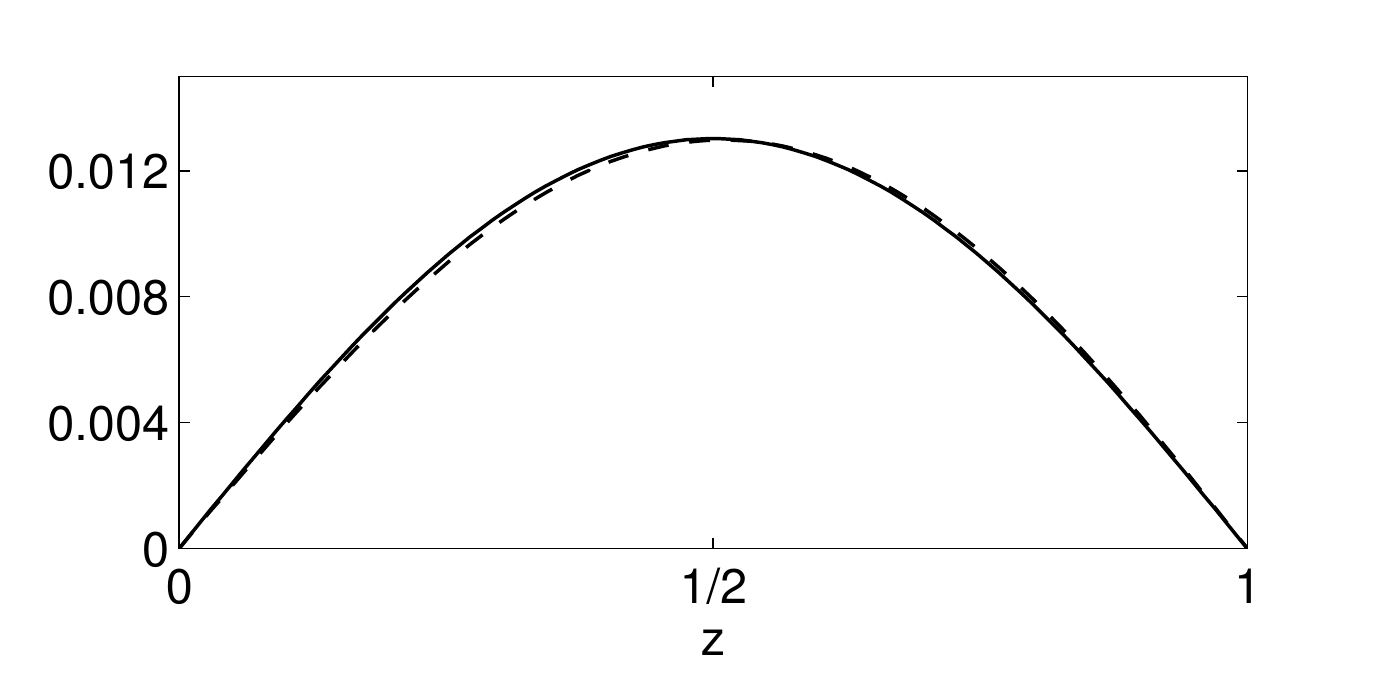}
\caption{The polynomials $P(z)$ ($\solidrule$) and $Q_{\gamma}(z)$ ($\dottedrule$), where the latter are evaluated for the optimal coupling parameter $\gamma^*$. Top and bottom boundary conditions on the velocity are (a) both no-slip, (b) free-slip only at the top, (c) free-slip only at the bottom, and (d) both free-slip.}
\label{fig: poly}
\end{figure}

Figure \ref{fig: poly} shows that when the energy (\ref{eq: E_gamma}) is defined using the optimal coupling parameter $\gamma^*$ (cf.\ table \ref{tab: R_E}), each fifth-order $Q_\gamma(z)$ closely approximates the corresponding fourth-order $P(z)$. In the $k^2\to0$ limit, the linear stability eigenmode has vertical velocity proportional to $P(z)$, the energy stability eigenmode has vertical velocity proportional to $Q_\gamma(z)$, and both eigenmodes have constant temperature. Thus, the closeness of the optimal $Q_\gamma(z)$ to $P(z)$ means that the long-wavelength solutions to the linear and energy stability eigenproblems are similar not only in their generalized eigenvalues, $R_L$ and $R_E$, but also in their eigenfunctions.

\bibliographystyle{jfm}
\bibliography{library}

\end{document}